\documentclass[aps,prl,superscriptaddress,unsortedaddress,twocolumn]{revtex4}
\usepackage{amsmath,epsfig,mathrsfs,graphicx,amssymb}

\begin{document}
\title{Nonsymmetrized Correlations in Quantum Noninvasive Measurements}
\author{Adam Bednorz}
\affiliation{Faculty of Physics, University of Warsaw ul. Ho\.za 69, PL00-681 Warsaw, Poland}
\author{Christoph Bruder}
\affiliation{Department of Physics, University of Basel,  Klingelbergstrasse 82, CH-4056 Basel, Switzerland}
\author{Bertrand Reulet}
\affiliation{D{\'e}partement de Physique, Universit{\'e} de Sherbrooke, Sherbrooke, QC, J1K2R1,Canada}
\author{Wolfgang Belzig}
\affiliation{Fachbereich Physik, Universit{\"a}t Konstanz, D-78457 Konstanz, Germany}
\date{\today}

\begin{abstract}
A long-standing problem in quantum mesoscopic physics is which
operator order corresponds to noise expressions like $\langle
I(-\omega)I(\omega)\rangle$, where $I(\omega)$ is the measured current
at frequency $\omega$. Symmetrized order describes a classical
measurement while nonsymmetrized order corresponds to a quantum
detector, e.g., one sensitive to either emission or absorption of
photons. We show that both order schemes can be embedded in quantum
weak-measurement theory taking into account 
measurements with memory, characterized by a memory function which is independent of a
particular experimental detection scheme. We discuss the resulting quasiprobabilities for different detector temperatures and how their negativity can be tested on the level of second-order correlation functions already. Experimentally, this negativity can be related to the squeezing of the
many-body state of the transported electrons in an ac-driven tunnel
junction.
\end{abstract}
\maketitle

Although quantum measurement theory has been based on the projection postulate
\cite{neumann}, nowadays it includes generalized schemes
based on auxiliary detectors, described mathematically by positive
operator-valued measures (POVM) \cite{povm}.  To specify a POVM
requires arguments based on physical considerations such as detector
efficiency, or the assumption of thermal equilibrium. A real physical
interaction generally leads to backaction on the system to be measured,
which makes the interpretation of measurements difficult.
Hence, all detection schemes are in general invasive as the measured
system is perturbed. The disturbance is strongest for projective
measurements, as the information in the measurement basis is
completely erased. In contrast, other POVM schemes can be much less
disturbing, as is often the case in experiments \cite{exper,exper2,ex3}.

To avoid invasiveness, Aharonov, Albert and Vaidman \cite{weak}
studied the limit of a {\it weak measurement}, in which the system is coupled
so weakly to the detector that it remains almost untouched. The price
to pay is a large detection noise, which is however completely
independent of the system. The gain is that other measurements on a
non-compatible observable can be performed. After the subtraction of
the detector noise, the statistics of the measurements has a
well-defined limit for vanishing coupling, which for incompatible
observables turns out to be described by a quasiprobability and not a
real probability distribution \cite{bednorz:prl10,abz}.

The most common weak-measurement theories assume that the
system-detector interaction is instantaneous
\cite{wvcond,aud,muk,tsang,schleich,jordan1}.  Such a Markovian
measurement scheme is relevant for many experiments \cite{exper} and
corresponds to the symmetrized order of operators:
$\langle I(-\omega)I(\omega)\rangle\to
\langle\hat{I}(-\omega)\hat{I}(\omega)+
\hat{I}(\omega)\hat{I}(-\omega)\rangle/2$, 
where quantum expectation values are defined as $\langle \hat{X}\rangle
=\mathrm{Tr}\hat{X}\hat{\rho}$ for an initial state
$\hat{\rho}$. Here, $\hat I(\omega)=\int \mathrm{d}t\; \hat I(t)e^{i\omega t}$
is the Fourier transform of the time-dependent current $\hat I(t)$ in the
Heisenberg picture. However, certain experiments are well described by 
nonsymmetrized
correlators like $\langle\hat{I}(-\omega)\hat{I}(\omega)\rangle$.
For $\omega\geq 0$ 
\cite{aguado,exper2,Lesovik,Gavish,clerk,bednorz:prb10,beenscho}
this corresponds to noise emitted by the system (emission noise) 
which is measured e.g. by an absorptive photo detector. These experiments
clearly lie beyond the scope of Markovian weak-measurement theory.

In this Letter, we formulate a general theory of weak detection which
allows for the description of nonsymmetrized correlators.  
We show that emission noise ($\omega\geq 0$) and absorption 
noise ($\omega< 0$) appear naturally if one
takes into account measurements with \emph{memory}. In fact, non-Markovian
weak measurements follow just from a few natural assumptions imposed
on the POVM in the limit of weak coupling. The results are independent
of a particular experimental realization and depend only on a single
memory function. By further requiring that no information transfer
occurs in thermal equilibrium the scheme is fixed uniquely and
contains only the detector temperature as a parameter. Varying the
detector temperature interpolates between emission and absorption
measurements. As our scheme is independent of other
properties of the detector, it applies to circuit QED,
mesoscopic current measurements and quantum optical systems equally
well.  Interestingly, applied
to a simple harmonic oscillator, the correlation functions in this scheme are consistent with
the Glauber-Sudarshan $P$-function \cite{glauber} known from quantum optics for absorption detectors.  Contrary to the
instantaneous measurements, the non-Markovian scheme can violate weak
positivity \cite{bednorz:prb10}.  To test it, we propose a measurement
of photon-assisted current-fluctuations, which are shown to violate a
Cauchy-Schwarz type inequality, proving the negative quasiprobability of the statistics after deconvolution of the detection noise. Identifying the finite-frequency
current operators with quadratures in analogy to quantum optics, we
show that the thus created non-equilibrium state of the current is
squeezed and therefore has essentially non-classical correlations.

We start by developing a general framework of weak quantum measurement based upon the POVM formalism including non-Markovian features.
We consider a set of $n$ \emph{independent} detectors continuously recording $n$ time-dependent signals $a_j(t)$ for $j=1,\ldots,n$. Each detector is related to an observable $\hat{A}_j$. For example, $n$ ammeters are inserted in a complex circuit: $a_j(t)$ is the recorded current in the branch $j$ and $\hat I_j(t)$ the current operator in that branch. Note that in general $\hat A_j(t)$ and $\hat A_k(t')$ do not commute even if $j\neq k$ since $\hat A_j$ and $\hat A_k$ may not commute with the Hamiltonian.
We want to relate classical correlators of measured quantities like $\langle a_1(t_1)\cdots a_n(t_n)\rangle$ to their equivalent for weak quantum measurements $\langle\cdots\rangle_w$.
These should involve \emph{linear} correlators of the $\hat{A}_j$, which can be taken at different times to allow
for memory effects of the detectors, while preserving \emph{causality}. The requirements of linearity and causality are fulfilled by replacing $a_j(t)$ in the correlator by  a superoperator $\int dt' \check{A}_j^{t-t'}(t')$ and perform time order, i.e.:
\begin{eqnarray}
&&\langle a_1(t_1)\cdots a_n(t_n)\rangle_w=\nonumber\\
&&\mathrm{Tr}\int \mathrm{d}^nt'\:\mathcal T\left[\check{A}_n^{t_n-t'_n}(t'_n)
\cdots\check{A}_1^{t_1-t'_1}(t'_1)\right]\hat\rho\: .
\label{avgabs}
\end{eqnarray}
Here $\mathcal T$ denotes time order with respect to the arguments in
brackets, $\hat\rho$ is the density matrix, and $\check{A}$ are superoperators defined as:
\begin{equation}
\check{A}_j^{t-t'}(t')
=g_j(t-t')\check{A}^c_{j}(t')+f_j(t-t')\check{A}^q_j(t')/2\:.
\end{equation}
The superoperators $\check{A}_j^{c/q}$ \cite{zwanzig} act on any
operator $\hat{X}$ like an anticommutator/commutator:
$\check{A}_j^c\hat{X}=\{\hat{A}_j,\hat{X}\}/2$ and
$\check{A}_j^q\hat{X}=[\hat{A}_j,\hat{X}]/i\hbar$.  
In the above expressions
we assumed for simplicity that the detectors are in a stationary
state so that only time differences $t_j-t_j'$ matter. 

We will also assume that the average of single measurements coincides with the usual average for projective measurements, i.e. $\langle a_j(t)\rangle_w=\langle\hat{A}_j(t)\rangle$. 
This implies $g_j(t-t')=\delta(t-t')$. Other choices of $g$ simply mimic the effect of classical frequency filters. 
Thus the only freedom left is the
choice of the real function $f_j$ that multiplies $\check{A}_j^q$. Note that $f_j(t)$
can be non-zero for $t>0$ without violating causality, since it is
accompanied by $\check{A}_j^q$ and only future measurements are
affected. For the last measurement, future effects disappear because
the leftmost $\check{A}^q$ vanishes under the trace in Eq. (\ref{avgabs}).
For simplicity, we will assume a single $f=f_j$, independent of $j$.
The limit $f=0$ corresponds to the Markovian case.

Now we want to show that correlations obeying these requirements can
be obtained from the general quantum measurement formalism.  Based on
Kraus operators $\hat{K}$ \cite{kraus}, the probability distribution
of the measurement results is $\rho=\langle\check{K}\rangle$ for
$\check{K}\hat{X}=\hat{K}\hat{X}\hat{K}^\dag$, where the only
condition on $\hat{K}$ is that the outcome probability is normalized
regardless of the input state $\hat{\rho}$. Here we need $\hat{K}$ to
be time-dependent. In general, we assume that $\hat{K}[\hat{A},a]$ is
a functional of the whole time history of observables $\hat A(t)$ and
outcomes $a(t)$. We shall assume that the functional $\hat{K}$ is
stationary so it depends only on relative time arguments.

The essential step to satisfy Eq.~(\ref{avgabs}) is to take the limit
$\hat{K}\sim \hat{1}$
which corresponds to a noninvasive measurement.
This can be obtained from an arbitrary initial POVM by rescaling
$\hat{K}[\hat{A},a]\to \hat{K}_\eta =C(\eta)\hat{K}[\eta\hat{A},\eta
  a]$ with $\eta\to 0$, which defines
$\rho_\eta=\langle\check{K}_\eta\rangle$. Here $C(\eta)$ is a
normalization factor.

The desired correlation function (\ref{avgabs}) can be derived by the
following limiting procedure for an almost general POVM,
\begin{equation}
\langle a_1(t_1)\cdots a_n(t_n)\rangle_w=
\lim_{\eta\to 0}\langle a_1(t_1)\cdots a_n(t_n)\rangle_\eta\:,
\end{equation}
where the average on the right-hand side is with respect to
$\rho_\eta$.  We assume the absence of internal correlations between
different detectors, namely $\hat{K}[\hat{A},a]=\mathcal
T\prod_j\hat{K}[\hat{A}_j,a_j]$, where $\mathcal T$ applies to the
time arguments of $\hat{A}$.

Expanding
$\hat{K}[\hat A,a]/k[a]=1+\int \mathrm{d}t' F[a,t']\hat A(t') +
\mathcal O(\hat A^{2})$,
we find, up to $\mathcal O(\hat{A}^{2})$,
\begin{equation}
\check{K}/|k[a]|^2\simeq 1
+\int \mathrm{d}t'\;\left(2\mathrm{Re}F\check{A}^c(t')
-\hbar\mathrm{Im}F\check{A}^q(t')\right)\:.
\label{main_result}
\end{equation}
Here, $|k[a]|^2$ is a functional probability of time-resolved outcomes independent of
the properties of the system which represents the detection noise.
As we want the measurement to be noninvasive to lowest order, we
impose the condition that $\int F[a,t']|k[a]|^2\mathrm{D}a$ vanishes;
$\mathrm{D}a$ is the functional measure.  Our conditions imply that
$\int 2a(t)\mathrm{Re}F[a,t']|k[a]|^2\mathrm{D}a=\delta(t-t')$, and
we get $f(t-t')=-\int 2a(t)\hbar\mathrm{Im}
F[a,t']|k[a]|^2\mathrm{D}a$.  Thus, the most general weak Kraus
operator takes the form given in Eq.~(\ref{main_result}), which is
our main result.  A particular Gaussian example of a POVM realizing
this scheme is presented in the Supplemental Material A.  We emphasize that our measurement scheme is not limited to any particular model of a detector, rather is captures generic properties of a general weakly invasive detector, whose property is encoded in the choice of the real function $f(t)$.

To discuss the consequences of different forms of $f$, we now calculate
the noise spectral density,
\begin{equation}
S_{ab}(\omega)=\int\mathrm{d}t\;e^{i\omega t}\langle a(t)b(0)\rangle_w\:.
\label{noii}
\end{equation}
An important special case is a system in a thermal equilibrium state,
$\hat{\rho}\sim\exp(-\hat{H}/k_BT)$. We further assume that the
averages of $\hat{A}$ and $\hat{B}$ vanish.  If the detector
temperature $T_d$ is equal to $T$ and in the absence of other
nonequilibrium effects (like a bias voltage, or special initial
conditions), we expect that no information transfer from the system to
the detector occurs, i.e., that $S_{ab}(\omega)=0$.  This requirement
leads to a necessary condition on the form of $f$
(see Supplemental Material B):
$f(\omega)=i\hbar(2n_B(\omega)-1)=i\hbar\coth(\hbar\omega/2k_BT_d)$, where
$n_B(\omega)$ is the Bose distribution at temperature $T_d$.
Equivalently, $f(t)=|k_BT_d|\coth(\pi tk_BT_d/\hbar)$ (at zero
temperature $f(t)=\hbar/\pi t$). 
We use the name {\it equilibrium order} for this special choice of $f$. The zero temperature case has been also called \emph{time-normal} \cite{plimak}. It is relevant
for experimental situations like in \cite{exper2} and consistent with the
quantum tape \cite{bednorz:prb10} or photodetection model
\cite{beenscho} if the temperature of the tape (or the photons) is
$T_d$. 

The necessary form of $f$ is also sufficient.  Indeed, the property
$S_{ab}(\omega)=0$ follows from the fluctuation-dissipation theorem
\cite{fdt} ($\langle \cdots \rangle_T$ denotes the equilibrium average)
\begin{equation}
 \int \mathrm{d}t\; e^{i\omega t}
  \langle\hat{A}(t)\hat{B}(0)\rangle_T
  =\int \mathrm{d}t\; e^{i\omega t+\hbar\omega/k_BT}
  \langle\hat{B}(0) \hat{A}(t)\rangle_T\:,
\end{equation}
because for an arbitrary stationary state we get
\begin{eqnarray}
&&S_{ab}(\omega)=\int \mathrm{d}t\; e^{i\omega t}
  \langle e^{\hbar\omega/2k_BT_d}\hat{B}(0)\hat{A}(t)\nonumber\\
    &&-e^{-\hbar\omega/2k_BT_d}\hat{A}(t)\hat{B}(0)
\rangle/\sinh(\hbar\omega/2k_BT_d)\:.
\label{fdt1}
\end{eqnarray}
For zero detector temperature, this reduces to 
\begin{equation}
\int e^{i\omega t}\mathrm{d}t\;\langle\theta(-\omega)\hat{A}(t)\hat{B}(0)+
\theta(\omega)\hat{B}(0)\hat{A}(t)\rangle\,,
\label{sem}
\end{equation}
which corresponds to emission noise for $\hat A=\hat B$ and $\omega>0$\cite{clerk}. 
Note that in that case (\ref{fdt1}) and (\ref{sem}) are even functions of $\omega$ but
the operators do not appear in symmetrized form. 
Thus, for $T_d \neq T$,
$S_{ab}(\omega)$ is in general not equal to zero and contains
information about the system. 
It is interesting to note that reversing the sign of
$f$ transforms $S_{ab}(\omega)$  into absorption noise for $\omega\leq 0$. Hence, measuring absorption noise requires a detector formally described by a negative temperature $T_d$ in $f$ and Eq.~(\ref{fdt1}).

It is interesting to note that for this special choice of $f$ the
higher-order fluctuations also vanish if
$\hat{\rho}\propto\exp(-\hat{H}/k_BT)$ and $T=T_d$.  We can write the
Fourier transform of (\ref{avgabs}) as 
\begin{equation}
\int \mathrm{d}^nt\;e^{i\sum_k\omega_kt_k}\mathrm{Tr}\mathcal
T\prod_k\sum_\pm\frac{\pm e^{\pm\hbar\omega_k/2k_BT}
\check{A}^\pm_k(t_k)}{2\sinh(\hbar\omega_k/2k_BT)}\hat{\rho}\:,
\label{avgabs1}
\end{equation}
with $\check{A}^+\hat{X}=\hat{A}\hat{X}$ and
$\check{A}^-\hat{X}=\hat{X}\hat{A}$.  Now, we can split
$\hat{\rho}=\hat{\rho}^{1/2}\hat{\rho}^{1/2}$, expand the above
expression as a sum of operator products and move one factor
$\hat{\rho}^{1/2}$ leftwards and the other rightwards so that they
meet again at the trace sign, which gives (\ref{avgabs}) in the form
\begin{eqnarray}
&&\int \mathrm{d}^nt\;e^{i\sum_k\omega_kt_k}
\mathrm{Tr}\hat{\rho}\mathcal
 T\prod_k\\
&&\sum_\pm\frac{\pm e^{\pm\hbar\omega_k/2k_BT}
\check{A}^\pm_k(t_k\mp i\hbar/2k_BT)}{2\sinh(\hbar\omega_k/2k_BT)}\:.
\nonumber
\end{eqnarray}
Shifting $t\to t\pm i\hbar/2k_BT$ and using Tr$\check A^q\ldots=0$ leads to
\begin{equation}
\int \mathrm{d}^nt\;e^{\sum_ki\omega_kt_k}\mathrm{Tr}\hat{\rho}\mathcal
 T\prod_k
\check{A}^q_k(t_k)/2i\sinh(\hbar\omega_k/2k_BT)=0\:.
\end{equation}
The vanishing of all correlations means that at any temperature
$T=T_d$ in equilibrium no information is transferred, not even in
higher-order correlators. This is a property of the memory function
$f$ in the idealized limit of our weak detection scheme. 
In the photoabsorption scheme at zero temperature, these properties
are intuitively clear \cite{beenscho}, because no photons are emitted.

Subtracting the (large) detection noise (by deconvoluting the $|k[a]|^2$ term in Eq. (\ref{main_result})),
the correlations defined in Eq.~(\ref{avgabs}) can be described by a quasiprobability.
Surprisingly, the equilibrium order differs qualitatively from the
symmetrized one, when one considers weak positivity, i.e. second-order correlations 
can be obtained from a positive probability \cite{abz,prb11}.
The symmetrized correlation matrix $C_{ab}=\langle
ab\rangle=\langle\hat{A}\hat{B}+\hat{B}\hat{A}\rangle/2$ is positive
definite, and the Gaussian probability distribution $\propto
\exp(-\sum_{ab}C^{-1}_{ab}ab/2)$ reproduces all first and second-order
symmetrized quantum correlations. This is not the case for equilibrium
order. If $T_d>T$ the autocorrelation is negative meaning that also the underlying 
quasiprobability is negative. E.g.~ at $T=0$ and $\omega>0$ Eq.~(7) gives 
$\langle a(-\omega) a(\omega)\rangle_w =-\langle \hat{A}(\omega)\hat{A}(-\omega)\rangle
e^{-\hbar\omega/2k_BT_d}/2\sinh(\hbar\omega/2k_BT_d)$.
In a stationary situation and for $T_d=0$, the weak positivity
holds, because the correlation matrix (\ref{sem}) is positive
definite. However, it can be violated in nonstationary situations.
This can be demonstrated using a two level system with the Hamiltonian
$\hat{H}=\hbar\Omega\hat{\sigma}_z/2$, with observables
$\hat{A}=\hat B=\hat{\sigma}_x+\hat{\sigma}_z$ and the initial state
$\hat\rho(0)=(\hat{1}+\hat{\sigma}_y)/2$.
By direct calculation (Supplemental
Material C) we find $\langle a(0)b(0)\rangle_w=-(2/\pi)\ln \Omega
t_\infty$, where $t_\infty$ is a cutoff set by
intrinsic decoherence or detector backaction. Since the observables measured by the two detectors are the same, $\langle a(0)b(0)\rangle_w=\langle a^2(0)\rangle_w$ and weak
positivity is obviously violated. 

There is an interesting connection between equilibrium order and
the Glauber-Sudarshan $P$ function \cite{glauber}.  Let us take the
harmonic oscillator $\hat{H}=\hbar\Omega\left(\hat{p}^2+\hat{x}^2\right)/2$, with
$[\hat{x},\hat{p}]=i$ and $\Omega>0$, and consider correlations with respect to
the quasiprobability (\ref{avgabs}) with antisymmetric $f(t)$. In this
case the time order is irrelevant as shown in Supplemental
Material D.  Let us define $\check{A}(t)=\check{A}^c(t)+\int
dt'\;f(t-t')\check{A}^q(t')/2$. The evolution of $\check{x}(t)$ and
$\check{p}(t)$ is just classical,
\begin{equation}
\label{eq:osci}
\check{x}(t)  = \check{x}\cos(\Omega t)+
\check{p}\sin(\Omega t),\;
\check{p}(t) =  \check{p}\cos(\Omega t)-\check{x}\sin(\Omega t)
\end{equation}
while  for $t=0$ (since $f(t)$ is antisymmetric and real) we have
$
\check{x}=\check{x}^c+i\check{p}^qf(\Omega)/2$, $\check{p}=\check{p}^c-i\check{x}^q f(\Omega)/2$.
Eq.~(\ref{eq:osci}) means that all correlators in this case undergo the classical time evolution.
We now define the ladder operators through
$\hat{a}^\dagger=(\hat{x}+ i\hat{p})/\sqrt{2}$
with $[\hat{a},\hat{a}^\dagger]=\hat{1}$.  This leads to $\check{a}=\check{a}^c- f(\Omega)\check{a}^{q}/2$ and $\check{a}^\dagger=\check{a}^{\dagger
  c} + f(\Omega)\check{a}^{\dagger q}/2$.  In the zero-temperature case,
$f(\Omega)=i\hbar$ (a perfect photodetector),
and defining $\alpha=(x+ip)/\sqrt{2}$ we get the single-time
quasiprobabilistic average $\langle\alpha^n\alpha^{\ast k}\rangle=
\mathrm{Tr}\hat{a}^n\hat{\rho}\hat{a}^{\dagger k}$.  On the other hand,
this is a property of the Glauber-Sudarshan function $P(\alpha)$, defined
by
$\hat{\rho}=\int\mathrm{d}^2\alpha\;P(\alpha)|\alpha\rangle\langle\alpha|$
for normalized coherent states
$\hat{a}|\alpha\rangle=\alpha|\alpha\rangle$,
$\langle\alpha|\alpha\rangle=1$ \cite{glauber}.  Since $\langle\alpha^n\alpha^{\ast k}\rangle_P=\int
\mathrm{d}^2\alpha\;\alpha^n\alpha^{\ast
  k}P(\alpha)=\mathrm{Tr}\hat{a}^n\hat{\rho}\hat{a}^{\dagger k}$, we
find that the quasiprobability for a zero-temperature detector is identical to $P(\alpha)$. It is interesting to note that
reversing the sign of $f$ leads to the Husimi-Kano $Q$ function instead of
$P$ \cite{povm}, while $f=0$ gives the Wigner function \cite{bednorz:prl10,bednorz:pra11}.

The fact that we obtain the $P$-function shows the deep connection
between the non-Markovian weak measurement formalism and the
quantum-optical detector theory.  One of the interesting consequences
is that zero-temperature equilibrium order is consistent with
photoabsorptive detection schemes, in which the $P$-function appears
naturally \cite{povm}.  It is also interesting to draw a link between
the violation of weak positivity in equilibrium order and the
properties of squeezed states. The ground state of a harmonic
oscillator fulfills $\langle\hat{x}^2\rangle=1/2$, which
corresponds to $\langle x^2\rangle_P=0$. A squeezed state can be such
that $\langle\hat{x}^2\rangle<1/2$, still minimizing the
Heisenberg uncertainty principle. This translates into a negative
variance of the position described by the (quasiprobability)
$P$-function, i.e. $\langle x^2\rangle_P<0$ \cite{squee} and is therefore
equivalent to a violation of weak positivity.

Let us now consider how our results apply to the case of current
fluctuations in mesoscopic conductors. The quantum description of the
noise in the junction, $S_I(\omega)=\int \mathrm{d}t\;e^{i\omega t}
\langle \delta I(t)\delta I(0)\rangle$, where $\delta
I(t)=I(t)-\langle I(t) \rangle$, will depend on the choice of $f$ in
(\ref{avgabs}).  For $f=0$, we get symmetrized noise $S^s_I=G\hbar
\sum_\pm w(\omega\pm eV/\hbar,T)/2$, where $G$ is the conductance, $V$
is the constant bias voltage and
$w(\alpha,T)=\alpha\coth(\hbar\alpha/k_BT)$ \cite{bednorz:prb10}.  For f given by equilibrium order with an arbitrary $T_d$, we obtain $S_I=S^s_I-G\hbar
w(\omega,T_d)$. Hence, the detection schemes differ by a term that is
independent of the voltage and the temperature of the system, making
it impossible to detect non-classicality in this scheme.

\begin{figure}
\includegraphics[scale=0.9]{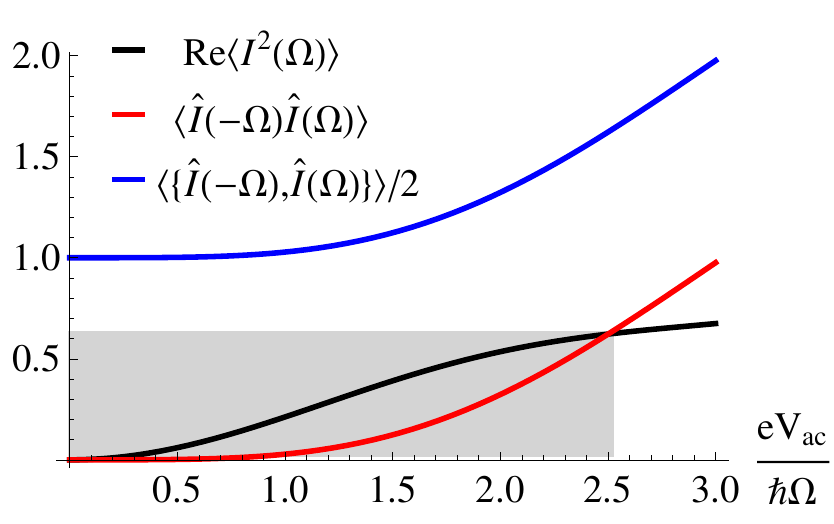}
\caption{(color online) Quantum correlation
  functions (in units of $2\pi G\hbar\Omega t_0$) for a tunnel junction at 
  zero temperature $T=0$.  The emission noise 
  $\langle \hat I(-\Omega)\hat I(\Omega)\rangle$ (red line) violates the
  classical inequality (\ref{eq:currentineq}) for a certain range of
  $eV_{ac}$ (shaded region). This violation is equivalent
  to the squeezing condition (\ref{eq:squeezing}) for the symmetrized
  noise $\langle\{\hat I(-\Omega),\hat I(\Omega)\}\rangle/2$ (blue line).}
\label{ff1}
\end{figure}

An experimentally feasible test of squeezing and violation of weak
positivity is possible using a coherent conductor (e.g. a tunnel
junction for the sake of simplicity) subject to an AC voltage bias
$V(t)=V_{ac}\cos\Omega t$\cite{gabelli:12}. Consider the classical inequality
\begin{equation}
	\label{eq:currentineq}
	|\delta I(\omega)-\delta I(-\omega)|^2\geq0 \Rightarrow
	\langle |\delta I(\omega)|^2\rangle\geq \mathrm{Re}\langle\delta I^2(\omega)\rangle\,.
\end{equation}
For symmetrized order one gets \cite{acs}
\begin{eqnarray}
&&\langle \{\delta\hat I(\omega),\delta\hat I(\omega')\}\rangle/2=
2\pi\hbar G\sum_m\delta(\omega+\omega'-2m\Omega)
\nonumber\\
&&\sum_n J_n(eV_{ac}/\hbar\Omega)
J_{n-2m}(eV_{ac}/\hbar\Omega) w(\omega-n\Omega)\,,
\end{eqnarray}
where $J_n$ are the Bessel functions.  In the case of equilibrium
order at $T_d=0$ one only has to subtract $2\pi\hbar
G|\omega|\delta(\omega+\omega')$ from the above result. As shown in
Fig.~\ref{ff1}, the classical inequality is violated for
$\omega=\Omega$ in a certain range of $eV_{ac}/\hbar\Omega$, but only
in equilibrium order. This can be reinterpreted in terms of the
existence of squeezing in the quantum shot noise : consider the two
quadratures associated with the finite-frequency current operator:
$\hat{A}=i[\delta\hat{I}(\omega)-\delta\hat{I}(-\omega)]/2$ and
$\hat{B}=[\delta\hat{I}(\omega)+\delta\hat{I}(-\omega)]/2$. Using
$\langle[\hat{I}(\omega),\hat{I}(-\omega)]\rangle=2t_0G\hbar\omega$,
we find \cite{gyi} $\langle[\hat{A},\hat{B}]\rangle= it_0G\hbar\omega$ 
(with the total detection time $t_0$). Thus the squeezing condition \cite{povm},
\begin{equation}
	\label{eq:squeezing}
	\langle\hat{A}^2\rangle<|\langle[\hat{A},\hat{B}]\rangle|/2\,,
\end{equation}
is related to the violation of weak positivity, $\langle
A^2\rangle_w<0$ in equilibrium order with $T_d=0$ and allows to
violate Eq.~(\ref{eq:currentineq}). Hence, according to
Fig. \ref{ff1}, quantum shot noise with AC-driving creates current
states, which resemble squeezed light for a certain range of the
AC-voltage.

In conclusion, we have presented a theory of a generic
weak-measurement scheme that includes emission noise.  It requires a
non-Markovian POVM with a specially chosen memory function $f$, which
has no analog in the Markovian picture. The scheme is consistent with
the absence of information flow between system and detector in
equilibrium at a given temperature.  Hence any detection requires a
nonequilibrium situation. Another direct consequence is that even the
simple Markovian detection process must involve a nonequilibrium
detector state.  Finally, nonsymmetrized order leads to a violation
of weak positivity, which can be tested experimentally by violation of
suitable inequalities, equivalent to the squeezing condition in some
cases.

\emph{Note added:} Our 
prediction about AC-driven squeezing has been recently confirmed experimentally \cite{glr:13}.

We acknowledge useful discussions with A. Klenner, L. Plimak and
J. Gabelli. This work was financially supported by the Swiss SNF and
the NCCR Quantum Science and Technology (CB), the DFG through SFB 767
and SP 1285 (AB and WB), the Polish Ministry of Science grant IP2011
002371 (AB), and the Canada Excellence Research Chairs program (BR).

\onecolumngrid
\newpage
\vspace{5cm}

\begin{center}
\large\bf Supplemental Material
\end{center}
\vspace{1cm}

\section{A. Gaussian non-Markovian POVM}
\renewcommand{\theequation}{A.\arabic{equation}}
\setcounter{equation}{0}

An example of a POVM leading to a non-Markovian weak measurement is
based on the Gaussian detector prepared in the initial state (wavefunction) $\phi(x)\propto\exp(-x^2)$,
interacting with the system by the time-dependent Hamiltonian (in the interaction picture)
$H_I(t')=(\delta(t-t')p+2f(t-t')x)\hat{A}(t')$. The momentum $p$ makes the shift $x\to x-\hat{A}(t)$.
For the measurement of $a(t)=x$ we get
the Gaussian Kraus operator
\begin{equation}
\hat{K}[\hat{A},a]\propto\mathcal T \exp\left[-(\hat{A}(t)-a(t))^2-
\int \mathrm{d}t'\;2if(t-t')(a(t)-\hat{A}(t)\theta(t-t'))\hat{A}(t')/\hbar\right]\,.
\label{keq}
\end{equation}
Here, the first term in the exponent is the Markovian part, while the
second term describes the non-Markovian measurement process including
a fixed but arbitrary real function $f(t)$, characterizing the memory
effect. The Heaviside function $\theta$ follows from the fact that $p$ shifts the phase for $t'<t$ and ensures the
normalization of the Kraus operator.  
By comparing with (\ref{main_result}), we get $|k[a]|^2=\sqrt{2/\pi}e^{-2a^2}$
and $F[a,t']=2a(t)(\delta(t-t')-if(t-t')/\hbar)$, which in this special case are just usual functions.
Following the standard procedure
we find the Kraus superoperator in the form
\begin{eqnarray}
&&\check{K}[\hat{A},a]\propto\\
&&\mathcal T \exp\left[-2(\check{A}^c(t)-a(t))^2+
\frac{(\hbar\check{A}^q(t))^2}{2}
+ \int \mathrm{d}t'\;2f(t-t')(a(t)\check{A}^q(t')-
\theta(t-t')(\check{A}^c(t)\check{A}^q(t')+
\check{A}^q(t)\check{A}^c(t'))\right].\nonumber
\end{eqnarray}
To prove the normalization, $\int da
\langle\check{K}\rangle=1$, we perform the Gaussian integral
over $a$ (time order is no problem if kept up throughout the
calculation) and get
\begin{eqnarray}
&&\int da\;\check{K}=\mathcal T
\exp\left[(\hbar\check{A}^q(t))^2/2+
\int \mathrm{d}t' \theta(t'-t)
2f(t-t')\check{A}^q(t')\check{A}^c(t)\mathrm{d}t'\right.\nonumber\\
&&\left.-\theta(t-t')2f(t-t')\check{A}^q(t)\check{A}^c(t')\mathrm{d}t'
+ f(t-t')f(t-t'')\check{A}^q(t')
\check{A}^q(t'')\mathrm{d}t'\mathrm{d}t''/2\right],
\end{eqnarray}
where we have ordered properly $\check{A}^q(t')$ and $\check{A}^c(t)$.
In the power expansion, omitting the identity term, the leftmost
superoperator is always $\check{A}^q$. Since
$\mathrm{Tr}\check{A}^q\cdots=0$ we obtain $\int
\mathrm{d}a\;\langle\check{K}[\hat{A},a]\rangle=1$ or $\int
\mathrm{d}a\; \hat{K}^\dag\hat{K}=\hat{1}$.  In general, we define
$\hat{K}[\hat{A},a]$ for $n$ measurements as
$\hat{K}[\hat{A},a]=\mathcal T\prod_j \hat{K}[\hat{A}_j,a_j]$, taking $\hat{H}_I=\sum_j\hat{H}_{j,I}$. 
To get a weak measurement, we substitute $\hat{K}$ by $\hat{K}_\eta$ which is obtained by
replacing $\hat{H}_I\to\eta\hat{H}_I$ and measuring $a(t)=\eta x$.
Note
that putting $\hat{A}=0$ gives Gaussian white noise $\rho\propto
e^{-2a^2}$, which leads to large detection noise in the weak
limit, $\rho_\eta\propto e^{-2\eta^2 a^2}$, that has to be
subtracted/deconvoluted from the experimental data.

\section{B. Fixing the memory function $f$}
\renewcommand{\theequation}{B.\arabic{equation}}
\setcounter{equation}{0}

Since the detector function $f(\omega)$ should be system-independent
in thermal equilibrium, any system can be used to determine it. We
therefore consider a 2-level system
with $\hat{A}=\hat{B}=\hat{\sigma}_x$ and
$\hat{H}=\hbar\Omega\hat{\sigma}_z/2$.
The requirement $S(\omega)=0$ is equivalent to
\begin{equation}\mathrm{Re}
\int_{-\infty}^0 e^{i\omega t}\mathrm{d}t\;\langle(1+if(\omega)/\hbar)\hat{\sigma}_x(t)\hat{\sigma}_x(0)+
(1-if(\omega)/\hbar)\hat{\sigma}_x(0)\hat{\sigma}_x(t)\rangle=0\:.
\label{b1}
\end{equation}
The equilibrium state reads
$\hat{\rho}=(\hat{1}-\hat{\sigma}_z\tanh(\hbar\Omega/2k_BT))/2$
and
\begin{equation}
\langle\hat{\sigma}_x(0)\hat{\sigma}_x(t)\rangle=\langle\hat{\sigma}_x(-t)\hat{\sigma}_x(0)\rangle=\cos(\Omega t)+i\tanh(\hbar\Omega/2k_BT )\sin(\Omega t)\:,
\end{equation}
and (\ref{b1}) leads to the requirement that
\begin{equation}
\mathrm{Re}\left[
	\frac{1}{\epsilon+i\omega+i\Omega} + \frac{1}{\epsilon+i\omega-i\Omega}
	-if(\omega)/\hbar
\left(\frac{1}{\epsilon+i\omega+i\Omega}
-\frac{1}{\epsilon+i\omega-i\Omega}\right)\tanh\left(\frac{\hbar\Omega}{2k_BT}\right)
\right]\label{ress}
\end{equation}
vanishes for $\epsilon\to 0_+$.
Since $\frac{1}{\epsilon+ix}=\frac{1}{ix}+\pi\delta(x)$, we can ignore
the delta function for $\omega\neq \pm\Omega$, and the vanishing of
(\ref{ress}) reduces to
$
\mathrm{Re}f(\omega)=0\:.
$
As $f$ must be independent of the system, we are free to choose
$\Omega$, $\mathrm{Re}f(\omega)=0$ must hold for all $\omega$,
including $\omega=\pm\Omega$.  So $f$ is purely imaginary,
and (\ref{ress}) reads
\begin{equation}
 \delta(\omega+\Omega) + \delta(\omega-\Omega)
 	+(\delta(\omega+\Omega)
 -\delta(\omega-\Omega))\tanh\left(\frac{\hbar\Omega}{2k_BT}\right)\mathrm{Im}f(\omega)/\hbar\:.
 \end{equation}
Therefore,
$1\mp\mathrm{Im}f(\omega)\tanh(\hbar\Omega/2k_BT)/\hbar=0$ at $\omega=\pm\Omega $,
and
$f(\omega)=i\hbar\coth(\hbar\omega/2k_BT)$.

\section{C. Violation of weak positivity}
\renewcommand{\theequation}{C.\arabic{equation}}
\setcounter{equation}{0}

From (\ref{avgabs}) we find for $f(t)=\hbar/\pi t$
\begin{equation}
\langle a(t)b(s)\rangle_w=\left\langle\{\hat{A}(t),\hat{B}(s)\}/2+
\int_{-\infty}^s i\mathrm{d}t'[\hat{A}(t'),\hat{B}(s)]/2\pi(t-t')+
\int_{-\infty}^t i\mathrm{d}s'[\hat{B}(s'),\hat{A}(t)]/2\pi(s-s')
\right\rangle\:.
\end{equation}
For $\hat{H}=\hbar\Omega\sigma_z/2$ and
$\hat{A}=\hat{B}=\hat{\sigma}_x+\hat{\sigma}_z$ we find
$\hat{A}(t)=\hat{\sigma}_x\cos\Omega t-
\hat{\sigma}_y\sin\Omega t+\hat{\sigma}_z$
and
$i[\hat{A},\hat{A}(t)]/2=\hat{\sigma}_x\sin\Omega t-
\hat{\sigma}_z\sin\Omega t+\hat{\sigma}_y(\cos\Omega t-1)$.
Therefore, for $\hat{\rho}(0)=(\hat{1}+\hat{\sigma}_y)/2$ we get
\begin{equation}
\langle a^2\rangle=2+\frac{2}{\pi}\int_0^\infty
\mathrm{d}t\frac{\cos\Omega t-1}{t}\:.
\end{equation}
For small $t$ the integral is convergent but for large $t$ only
$\cos\Omega t/t$ converges.  The remaining integral $\int
\mathrm{d}t/t$ diverges logarithmically and one should put a cutoff
at $t_\infty$.

Certainly no experiment will record infinite correlations. The cutoff
$t_\infty$ is in practice bounded by the decoherence time of the
system and the measurement noise (which also diverges). The infinity
would occur only in the limit of zero measurement strength and a
perfect two-level system, which is impossible.

\section{D. Harmonic oscillator}
\renewcommand{\theequation}{D.\arabic{equation}}
\setcounter{equation}{0}
For the harmonic oscillator with
$\hat{H}=\hat{p}^2/2m+m\omega^2\hat{x}^2$ and $[\hat{x},\hat{p}]=i\hbar$,
we have
$[\hat{x}(t),\hat{p}(t')]=i\hbar\hat{1}\cos(\Omega(t-t'))$,
$[\hat{x}(t),\hat{x}(t')]=-i\hbar\hat{1}\sin(\Omega(t-t'))/m\Omega$,
$[\hat{p}(t),\hat{p}(t')]=-i\hbar\hat{1}\sin(\Omega(t-t'))m\Omega$, so
the commutator depends only on the difference $t-t'$, which applies
also to superoperators.
To see that the time order is irrelevant, let us take linear
functions $A$,$B$ of $x$ and $p$ and calculate
\begin{eqnarray}
&&\left(\check{A}^c(t)+\int dt'\;f(t-t')\check{A}^q(t')/2\right)
\left(\check{B}^c(s)+\int ds'\;f(s-s')\check{B}^q(s')/2\right)\nonumber\\
&&-\mathcal T\left(\check{A}^c(t)+\int dt'\;f(t-t')\check{A}^q(t')/2\right)
\left(\check{B}^c(s)+\int ds'\;f(s-s')\check{B}^q(s')\right)\\
&&=\int dt'\;f(t-t')\theta(s-t')[\check{B}^c(s),\check{A}^q(t')]/2+\int ds'\;f(s-s')\theta(t-s')[\check{B}^q(t),\check{A}^c(s')]/2\nonumber\\
&&=\int du\; \theta(u)
(f(u+t-s)[\check{B}^c(s),\check{A}^q(s-u)]
+f(s-t-u)[\check{B}^q(t+u),\check{A}^c(t)])/2\:.
\nonumber
\end{eqnarray}
The last expression vanishes because of the antisymmetry of $f$ and
the fact that the commutators depend only on the difference in time
arguments. The proof generalizes to multiple products because $[\check{B}^{c/q}(s),\check{A}^{c/q}(u)]$ is proportional to
identity superoperator for $A,B=x,p$.
\end{document}